\documentclass[final,5p,times,twocolumn]{elsarticle}
\usepackage{graphicx}
\usepackage{dcolumn}
\usepackage{braket}
\usepackage{bm}
\usepackage{mathrsfs}
\usepackage{xcolor}
\usepackage[utf8]{inputenc}
\usepackage{booktabs}
\usepackage{tabularx} 
\usepackage{amsmath,amsfonts,amssymb}

\usepackage[normalem]{ulem}  

\journal{Annals of Physics}

\begin{document}

\begin{frontmatter}

\title{Transient concurrence for copropagating entangled bosons and fermions}

\author[uabc]{M. Á. Terán-Cruz}
\author[uabc]{Roberto Romo}
\affiliation[uabc]{organization={Facultad de Ciencias, Universidad Aut\'onoma de Baja California},
            city={Ensenada},
            postcode={22800},
            state={Baja California},
            country={Mexico}}

\author[unam]{Gast\'on Garc\'{\i}a-Calder\'on}
\affiliation[unam]{organization={Instituto de F\'{\i}sica, Universidad Nacional Aut\'onoma de M\'exico},
            city={Ciudad de M\'exico},
            postcode={04510},
            country={Mexico}}

\begin{abstract}
The transient dynamics of copropagating entangled bosons and fermions remain an unexplored aspect of quantum mechanics. We investigate how entanglement manifests itself in the spatiotemporal evolution of the particles using a modified version of the quantum shutter model.
We derive a transient concurrence as a dynamical indicator of entanglement and demonstrate that it modulates the interference structure of the joint probability density, thereby revealing the spatial and temporal regions where probabilistic bunching and antibunching phenomena emerge. Furthermore, we derive analytical expressions revealing a structural connection between concurrence and the cosine modulation characteristic of Hanbury-Brown and Twiss (HBT) interference patterns. In the stationary limit, the Wootters concurrence is shown to coincide with the interferometric visibility of the resulting pattern.
This work establishes a structural bridge between entanglement signatures and interference phenomena in transient copropagating systems, providing a theoretical framework for exploring their dynamical interplay.
\end{abstract}


\end{frontmatter}

\textit{Introduction.}---Quantum entanglement has been the subject of intense research since the early days of quantum mechanics \cite{epr,bohr_1935,sch_1935}. Traditionally, attention has focused on scenarios in which entangled particles move away from each other, with special attention to spin entanglement. However, there is a less explored regime that could offer new  physical insights: entanglement in copropagation scenarios, where the quantum particles travel along the same direction. Landmark experiments such as the Hanbury Brown and Twiss effect (HBT) \cite{HBT_Effect} and the Hong-Ou-Mandel effect (HOM) \cite{HOM_Effect}, which have demonstrated quantum interference not only with photons but also with material particles \cite{Aspect2019}, illustrate the rich dynamics and potential applications of entanglement in configurations where particles coexist in proximity and 
maintain phase coherence while traveling together.
Fascinating experiments involving streams or beams of particles have revealed collective behaviors that are unequivocally linked to the bosonic or fermionic nature of the moving particles \cite{Yasuda_1996,Ensslin1999,Yamamoto1999,Schellekens_2005,Iannuzzi2006,Jeltes2007, Rosenberg2022, caylaPRL2020}. Furthermore, if the particles initially start from an entangled configuration, quantum correlations between them are expected to manifest in flight through visible interferences. 

Bunching and antibunching in bosonic and fermionic systems have also been examined in several theoretical studies that analyze joint detection probabilities in two-component states of identical particles \cite{MarchewkaGranot2014,Sancho2014,MarchewkaGranot2015,MousaviMiretArtes2020}. These works show that detection-level signatures depend sensitively on the spatial structure and modal content of the underlying wavefunctions. In analyses of systems whose single-particle wavefunctions contain zeros \cite{MarchewkaGranot2014}, it has been found that these nodes can invert the usual bosonic and fermionic behaviors, giving rise to regimes where bosons exhibit antibunching and fermions bunching. In studies of two-particle interference through a double-slit configuration \cite{Sancho2014}, multimode conditions have been identified under which bosons display detection patterns similar to those of distinguishable particles. In investigations of the decay of two identical particles released from quantum traps \cite{MarchewkaGranot2015}, escape scenarios have been reported in which bosons are more likely than fermions to emerge in opposite directions. Related effects have also been analyzed in dissipative diffraction settings \cite{MousaviMiretArtes2020}, where modifications to the wavefunction induced by environmental coupling alter the resulting correlation patterns. Taken together, these studies illustrate the diverse physical situations in which bunching and antibunching arise, emphasizing that simultaneous-detection correlations in identical-particle systems are governed by the detailed structure and evolution of the wavefunction.

Despite advances in the understanding of quantum entanglement in systems where particles are spatially separated, the study of their evolution in the time domain when two particles travel together represents a problem of fundamental interest in quantum mechanics. Our study addresses this question by explicitly analyzing the spatiotemporal manifestation of entanglement in copropagating particles and its impact on the joint probability density. For this purpose, a particularly suitable model is the Moshinsky quantum shutter \cite{PhysRev.88.625}.
This model offers exact analytical solutions with explicit spatial and temporal dependence, allowing us to visualize how the propagation of entangled particles evolves in space and time.

The characterization of entanglement in systems of indistinguishable particles has been the subject of intense conceptual  debate. As pointed out in~\cite{IeminiVianna2013}, this type of entanglement is subtler than in systems of distinguishable particles and has led to distinct approaches based on two complementary aspects: \emph{particle entanglement}, which refers to the impossibility of assigning a complete set of properties to each particle individually~\cite{GhirardiMarinatto2004,GhirardiMarinatto2005}, and \emph{mode entanglement}, which arises from coherent superpositions in the occupation of experimentally distinguishable modes~\cite{Zanardi2002, WisemanVaccaro2003, Benatti2020}. 
Recent work on continuous-variable entanglement~\cite{Swain2022} has explored
generalized frameworks that extend beyond standard discrete-variable approaches. These
continuous-variable formulations provide a broad theoretical background.
Here we adopt a different route, following the operational mode-entanglement
framework in which quantum correlations are accessed through \emph{projective
filtering} in the sense of Lin and Fisher~\cite{lin_fisher_2007}. This
measurement-based viewpoint is the one relevant to the dynamical scenario
studied in this work. This perspective underlies the notion of
\emph{detector-level entanglement} emphasized by Tichy et al.~\cite{Tichy2013},
where correlations are not tied to particle labels but emerge from observable
interference patterns and coincidences in the detection process.

In our study, we use analytical solutions to the time-dependent Schrödinger equation, $\Psi(x_1,x_2,t)$, within the Moshinsky quantum shutter model. Although the quantum state is defined over continuous variables, the analysis of quantum correlations is carried out after evaluating the wavefunction at specific points, in a procedure equivalent to projective filtering. This allows the continuous system to be mapped to a two-qubit state in the computational basis, as shown by Lin and Fisher~\cite{lin_fisher_2007}. This methodological choice naturally places our analysis within the framework of mode entanglement. In particular, our approach focuses on the transient quantum dynamics and on the time-dependent emergence of entanglement signatures, aspects that are central to our study.

In this work, we derive a transient concurrence as a dynamical indicator of entanglement, providing a framework to  visualize entanglement signatures through interference phenomena in transient quantum systems where the physical proximity and copropagation of the entangled components are essential. Our analysis reveals how this transient concurrence not only captures the phenomenon of diffraction in time \cite{PhysRev.88.625} but also modulates the interference correlation of the joint probability density, highlighting the regions where probabilistic bunching and antibunching phenomena emerge. Furthermore, in the long-time asymptotic limit, our transient concurrence converges towards the well-known Wootters concurrence~\cite{PhysRevLett.80.2245}.

\textit{Quantum shutter and quantum transients.}---We begin by reviewing the
quantum-shutter dynamics for a single coherent beam, establishing the
framework needed to describe the two-component shutter states used in our analysis.
The quantum shutter setup  was proposed by Moshinsky to discuss the transient regime he named \textit{diffraction in time} \cite{PhysRev.88.625}. It refers to the solution of the one-dimensional Schr\"odinger equation
\begin{equation}
i\hbar \partial/{\partial t \Psi(x,t)}=H\Psi(x,t),
\label{e1}
\end{equation}
with the initial condition at $t=0$,
\begin{equation}
\Psi(x,t=0)=\exp{(ik x)}\Theta(-x).
\label{e2}
\end{equation}
The above situation may be visualized as a quasi-monochromatic particle beam of energy $E=\hbar^2 k^2/2m$ that moves along the $x$-axis whose motion is interrupted by a shutter situated at $x=0$. If at $t=0$ the shutter is opened instantaneously, the solution \( \Psi(x,t) \) may be obtained in an exact analytical fashion, namely,
\begin{equation}
\Psi(x,t) = M(y),
\label{e6}
\end{equation}
where \( M(y) \), known as the Moshinsky or transient function, is defined as
\begin{align}
M(y) &\equiv \frac{i}{2\pi}\ \int_{-\infty}^\infty \frac{e^{ik'x-i\hbar k'^2 t/2m}}{k'-k}\,dk' \nonumber \\
     &= \frac{1}{2}{\rm e}^{imx^2/2\hbar t}w(iy),
\label{MoshinskyDef}
\end{align}with the argument \( y \) given by
\begin{equation}
y = {\rm e}^{-i\pi/4}\left ( \frac {m}{2\hbar t} \right )^{1/2}
\left [x- \frac {\hbar k}{m}t \right ],
\label{e9}
\end{equation}
and the function \( w(iy)=\exp \left(y^2\right) \operatorname{erfc}\left(y\right) \) is the Faddeyeva–Terent’ev function~\cite{faddeyeva61,abramowitzchap7}, for which efficient numerical routines are available~\cite{poppe90}.

Moshinsky showed that at asymptotically long times the free evolving solution (\ref{e6}) goes into the stationary form of the wave function, namely,
\begin{equation}
\Psi(x,t) \to e^{ikx}e^{-i\hbar k^2t/2m}.
\label{e11}
\end{equation}
Furthermore, he demonstrated that the ratio of the corresponding transient current density to the stationary current may be written in terms of Fresnel integrals in an identical fashion as for the intensity of light for the Fresnel diffraction by a straight edge except that the corresponding argument is now given by (\ref{e9}), which exhibits a dependence on time that led him to name this transient phenomenon as diffraction in time. Some decades later Dalibard and collaborators confirmed experimentally that diffraction in time is a transient phenomenon that exists in nature \cite{Dalibard96}. Thereafter this transient phenomenon has been confirmed by further experimental work and given origin to a good number of works, as reviewed in Ref. \cite{cgcm09}.

\textit{Quantum shutter for two-component states.}---
Here we consider the quantum–shutter dynamics for a system described by two
spatial degrees of freedom evolving under the Hamiltonian
$H=H_1+H_2$, where each $H_j$ acts on the coordinate $x_j$.
The initial state is taken as 
\begin{equation}
\Psi_0(x_1,x_2)= 
\chi \psi_{\alpha}(x_1) \psi_{\beta}(x_2) \pm 
\zeta \psi_{\beta}(x_1) \psi_{\alpha}(x_2),
\label{eq3.13}
\end{equation}
defined in the region $(x_1,x_2)<0$ and vanishing elsewhere.
The plus sign refers to the symmetric combination for bosons and the minus sign to the antisymmetric one for fermions. In the above expression, $\chi$  and $\zeta$ are complex coefficients, and 
$\psi_{q}(x_{j})\equiv \psi_{q}(x_{j},0)$, with $q=\alpha,\beta$, $\alpha \neq \beta$ and $j = 1,2$.

As is well known, for identical noninteracting particles, the Hamiltonian is symmetric under the permutation of the indices of the particles \cite{Messiah2014}, and therefore it commutes with the exchange operator. This guarantees that the symmetry of the initial state is preserved during the time evolution. Hence, the solution $\Psi(x_1,x_2,t)$ may be written as
\begin{equation}
\Psi(x_{1},x_{2},t) = \chi \Psi_{\alpha}(x_{1},t)\Psi_{\beta}(x_{2},t) \pm \zeta \Psi_{\beta}(x_{1},t)\Psi_{\alpha}(x_{2},t),
\label{eq3.16}
\end{equation}
where $\Psi_{q}(x_{j},t)\equiv \Psi(x_{j},q,t)$ are the solutions of the
one-dimensional Schrödinger equation for each spatial coordinate.
The global dynamics of the (anti)symmetrized two-component state is thus
determined by the individual wavefunctions $\Psi_{\alpha}(x_{j},t)$ and $\Psi_{\beta}(x_{j},t)$.

\textit{Transient concurrence.}---As is well known, the concurrence \( C \) of a stationary entangled pure state of the form
 \begin{equation}
\ket{\psi}=\xi\ket{01} \pm \eta \ket{10}
\label{Normalizedket}
\end{equation}
within the two-dimensional subspace defined by the basis \(\{ \ket{01}, \ket{10} \}\) is given by \cite{PhysRevLett.80.2245, walborn2006},
\begin{equation}
C(\psi)= 2|\xi\eta|=2 |\xi| \sqrt{1 - |\xi|^2},
\label{eq_2}
\end{equation}
with complex coefficients \(\xi\) and \(\eta\), satisfying the normalization condition \( |\xi|^2 + |\eta|^2 = 1 \). We shall now follow a prescription analogous to the one leading to the expression above, but adapted to our dynamical scenario so as to capture transient features arising from the dynamical co-propagation problem.

Let us consider a system described by two spatial degrees of freedom, with its
state written in the (anti)symmetrized two-component form
\begin{equation}
\Psi\left(x_1, x_2, t\right) = \xi\Phi_{A}\left(x_1, x_2,t\right)  \pm \eta\Phi_{B}\left(x_1,x_2, t\right).
\label{eq_5}
\end{equation}
The component states \(\Phi_{A}\) and \(\Phi_{B}\) are defined in terms of the one-dimensional solutions $\Psi_{q}(x_{j},t)$ as:
\begin{subequations}
\begin{align}
\Phi_{A}(x_1,x_2,t) &\equiv \Psi_\alpha\left(x_1, t\right) \Psi_\beta\left(x_2, t\right), \\
\Phi_{B}(x_1,x_2,t) &\equiv \Psi_\beta\left(x_1, t\right) \Psi_\alpha\left(x_2, t\right).
\end{align}
\label{eq_6}
\end{subequations}

The wavefunction in Eq.~(\ref{eq_5}) belongs to a continuous-variable Hilbert space of infinite dimension. In studies of entanglement for continuous-variable systems, a standard and well-established approach is to apply a procedure known as \emph{projective filtering}, where the quantum state is projected onto a discrete subspace associated with localized measurement outcomes~\cite{lin_fisher_2007}. This projection effectively maps the continuous-variable state to a finite-dimensional representation corresponding to a two-qubit form, thereby enabling the use of standard entanglement quantifiers such as concurrence. One practical implementation consists in evaluating the wavefunction at fixed spatial positions, say $x_1 = a$ and $x_2 = b$, which yields
\begin{equation}
\Psi(a,b, t) = \xi\Phi_{A}(a,b,t)  \pm \eta\Phi_{B}(a,b, t).
\label{filt_eq10}
\end{equation}
In many treatments, projective filtering is implemented using spatial windows of finite width (“imperfect detector” scheme), followed by renormalization via integration over the detection regions~\cite{lin_fisher_2007}. In our case, however, we adopt a pointlike (“ideal detector”) filtering approach without renormalization. This choice is consistent with the use of non-normalizable wavefunctions and with the fact that the resulting state vector, though unnormalized, retains all relevant information about the entanglement structure through the relative amplitudes of its components. This filtered wavefunction will now serve as the basis for constructing a discrete state vector using the computational basis. This assignment is not merely a notational convenience. As pointed out in Ref.~\cite{lin_fisher_2007}, there exists an isomorphic relation between the filtered amplitudes and the discrete qubit basis, 
establishing an algebraic correspondence with a two-qubit representation, which enables the construction of a concurrence-type indicator.

In order to introduce a concurrence-type indicator for probing entanglement signatures in this system, we will next map the quantum states
$\Psi_q(x_j,t)$ into the computational basis $\ket{0}$ and $\ket{1}$ \cite{nielsen},
\begin{align}
\ket{0} = \begin{pmatrix} 1 \\ 0 \end{pmatrix}, \quad \ket{1} = \begin{pmatrix} 0 \\ 1 \end{pmatrix},
\label{eq_7}
\end{align}
which provides a clear and precise way of describing quantum states as in the mathematical language of quantum information theory. We construct the tensor-product components by combining the one-dimensional
wavefunctions with the computational basis as follows:%
\begin{subequations}
\begin{align}
|\alpha\rangle_1 &\equiv\Psi_\alpha(a, t) \ket{0}, & |\beta\rangle_{1} &\equiv \Psi_\beta(a, t)\ket{1}, \label{eq3a} \\
|\alpha\rangle_2 &\equiv\Psi_\alpha(b, t)\ket{0}, & |\beta\rangle_2 &\equiv\Psi_\beta(b, t)\ket{1}. \label{eq3b}
\end{align}
\label{eq_8}
\end{subequations}
It is important to note that the computational basis is orthonormal $\langle l \mid m\rangle=\delta _{l m}$. Using the above, we can construct the system's vector by expressing it through the specific configurations of the particles as follows:
\begin{equation}
|\Psi\rangle=\xi|\alpha\rangle_1 \otimes|\beta\rangle_2\pm \eta|\beta\rangle_1 \otimes|\alpha\rangle_2,
\end{equation}
where \(|\Psi\rangle\) conforms to the definition of an entangled pure state \cite{nielsen,Horodecki2009,cohen2019quantum}, and hence it cannot be factorizable.
To make these configurations explicit, we express them in the computational
tensor-product basis:
\begin{subequations}
\begin{align}
|\alpha\rangle_1 \otimes |\beta\rangle_2 = \Phi_{A} \ket{01}, \\
|\beta\rangle_1 \otimes |\alpha\rangle_2 = \Phi_{B}\ket{10},
\end{align}
\label{eq_9}
\end{subequations}
where
\begin{equation}
\ket{01} = \begin{pmatrix} 0 \\ 1 \\ 0 \\ 0 \end{pmatrix}, \quad \ket{10} = \begin{pmatrix} 0 \\ 0 \\ 1 \\ 0 \end{pmatrix}.
\label{eq_10}
\end{equation}
Consequently, the state vector of the system is precisely given by 
\begin{equation}
|\Psi\rangle = \xi \Phi_{A}(a, b, t) \ket{01} \pm \eta \Phi_{B}(a, b, t) \ket{10}.
\label{eq_11}
\end{equation}

Following the algebraic structure of the Wootters prescription~\cite{PhysRevLett.80.2245}, we can define a \emph{transient concurrence} in terms of the state vector $|\Psi\rangle$ of the system as
\begin{equation}
\mathscr{C}(\Psi)=\left|\left\langle\Psi^* \mid \left(\hat{\sigma}_y \otimes \hat{\sigma}_y\right) \mid \Psi\right\rangle\right|.
\label{eq_12}
\end{equation}
Here, $\hat{\sigma}_y$ is a Pauli matrix, and hence the matrix form of the spin-flip operator is
\begin{equation}
\hat{\sigma}_y \otimes \hat{\sigma}_y = \begin{pmatrix}
0 & 0 & 0 & -1 \\
0 & 0 & 1 & 0 \\
0 & 1 & 0 & 0 \\
-1 & 0 & 0 & 0
\end{pmatrix}.
\label{eq_14}
\end{equation}
By performing the inner product indicated in (\ref{eq_12}), we obtain,
\begin{equation}
\mathscr{C}(\Psi) = 2|\xi\eta\Phi_{A}\Phi_{B}|.
\label{eq_17}
\end{equation}
Using (\ref{eq_2}) in the above expression, we find that the transient concurrence $\mathscr{C}$ we just derived above turns out to be equal to the Wootters concurrence modulated by a temporal factor, namely,
\begin{equation}
\begin{aligned}
\mathscr{C}(\Psi) = \underbrace{2|\xi| \sqrt{1 - |\xi|^2}}_{\text{Wootters concurrence}} 
\underbrace{|\Phi_{A}(a, b, t)\Phi_{B}(a, b, t)|}_{\text{temporal modulation}}
\end{aligned}
\label{eq_17}
\end{equation}
The quantity $\mathscr{C}(\Psi)$  represents a dynamical indicator describing how the underlying two-mode entanglement becomes locally manifested during propagation. It consists of the stationary concurrence $C(\psi)$ modulated by a spatiotemporal factor arising from the transient evolution of the system. In the long-time asymptotic limit, $\mathscr{C}(\Psi)$ smoothly converges to the standard Wootters concurrence $C(\psi)$.

\begin{figure*}
\centering
\includegraphics[width=4.8in,height=3.6in]{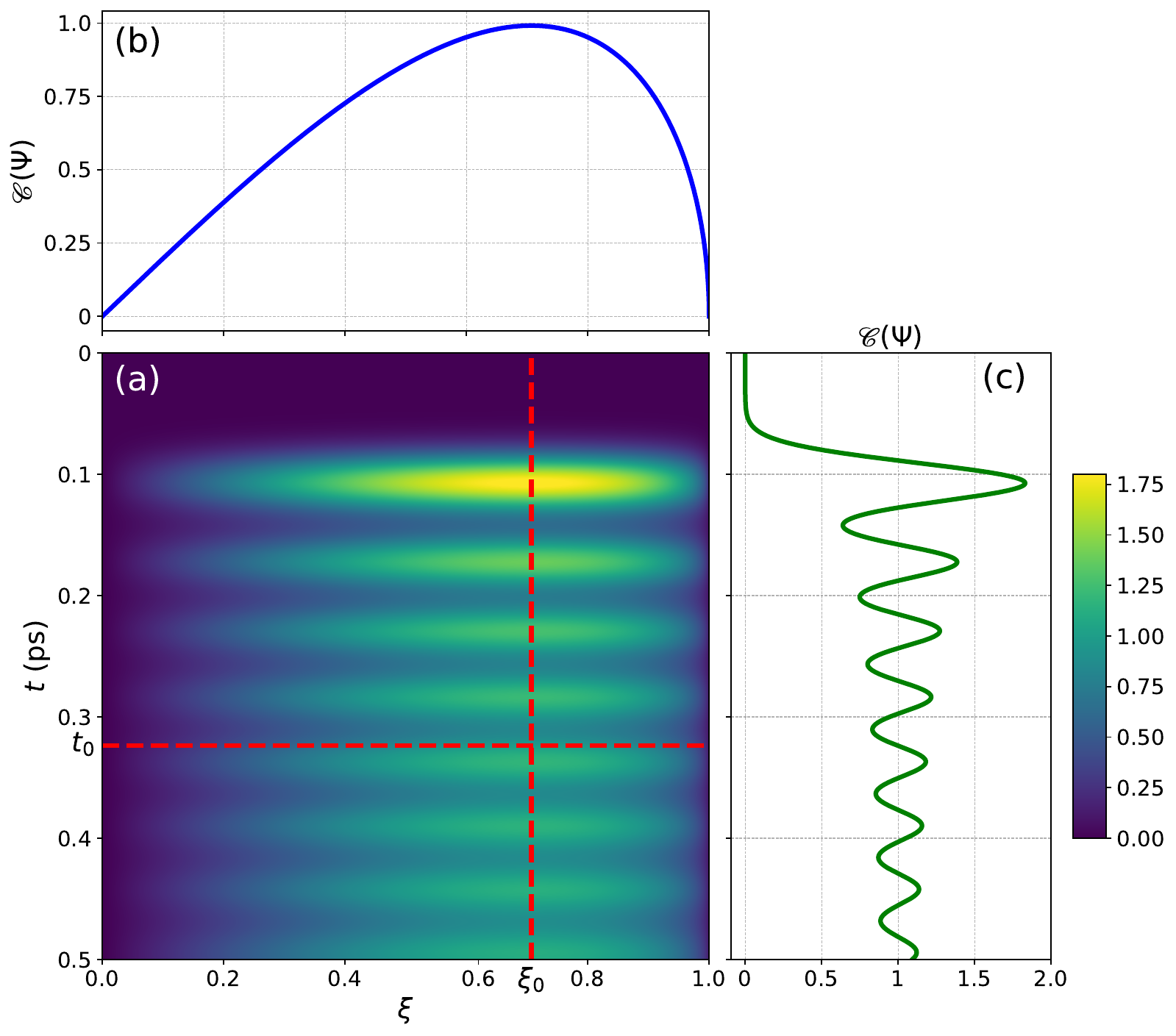}
\caption{\label{fig:wide}(a) Density plot  of the transient concurrence $\mathscr{C}(\Psi)$ calculated from Eq. (\ref{eq_17}) as a function of both the coefficient $\xi$ and time $t$ using the shutter model in the calculation of $|\Phi_{A}\Phi_{B}|$. As we can see in the upper panel (b), a section of the map at a fixed time, say $t_0=0.324$ ps (horizontal dashed red line), exhibits a curve with the characteristic shape of the concurrence of bipartite systems for a stationary entangled pure state \cite{walborn2006}. On the other hand, as we see in the right panel (c), a vertical cut in the map (vertical dashed red line) at a fixed value of $\xi$, say $\xi_0=1 / \sqrt{2}$, exhibits the emblematic profile of a diffraction in time pattern \cite{PhysRev.88.625}. Parameters: $a = 10.0$ nm, $b = 11.0$ nm,
$\alpha = 1.0 k_0$, $\beta = 1.01 k_0$,
with $k_0 = 1.449~\mathrm{nm}^{-1}$. The reference value of $k_0$ comes from
$k_0=\sqrt{2 m E_0}/\hbar$, where we chose the arbitrary reference energy
$E_0 = 0.08~\mathrm{eV}$ and the mass $m = m_e$ ($m_e$ is used here only as a reference mass).}
\label{Fig_1}
\end{figure*}

To evaluate $|\Phi_{A} \Phi_{B}|$, consider as an example the analytically solvable case of the quantum shutter model. The above consists of choosing $\Psi_{q}(x_{j},t)=M(y_{j,q})$ in Eqs. (\ref{eq_6}a) and (\ref{eq_6}b), where $y_{j,q}=\exp(-i \pi / 4 )\gamma_{j,q}$ with  $\gamma_{j,q}=(m/2\hbar t)^{1/2}(x_{j}-\hbar q t/m)$ ($q=\alpha,\beta$ and $x_j=a,b$), and then evaluate $\mathscr{C}$. In this case, $\Psi$ and $\mathscr{C}$ are of the form,
\begin{equation}
\Psi(a,b,t) = \xi M(y_{1,\alpha}) M(y_{2,\beta}) \pm \eta M(y_{1,\beta}) M(y_{2,\alpha}),
\label{Sol_Ms}
\end{equation}
\begin{equation}
\mathscr{C}(\Psi) = C(\psi)|M(y_{1,\alpha}) M(y_{2,\beta})M(y_{1,\beta}) M(y_{2,\alpha})|.
\label{concu_Ms}
\end{equation}

\textit{Visualizing the transient concurrence.}---Figure~\ref{Fig_1}(a) shows a density plot of the transient concurrence, illustrating how diffraction-in-time dynamics modulates entanglement signatures during propagation. In this map, $\mathscr{C}(\Psi)$ is plotted as a function of time and the parameter $\xi$. A horizontal cut at a fixed time $t_0$ yields a curve whose profile  resembles the Wootters concurrence $C(\psi)$ (upper panel), a curve experimentally verified by Walborn \textit{et al.}~\cite{walborn2006}.
Conversely, a 
vertical cut at a fixed value of $\xi$ (right panel) reveals temporal 
oscillations that evoke the diffraction-in-time pattern originally predicted 
by Moshinsky~\cite{PhysRev.88.625}, which was observed experimentally by 
Szriftgiser \textit{et al.}~\cite{Dalibard96}.

These graphs make explicit the interplay between two-mode entanglement, quantified by the Wootters concurrence, and the diffraction-in-time dynamics characteristic of quantum transients. In this way, the figure illustrates how a static entanglement measure becomes dynamically modulated during transient propagation. 
Note that as $t \rightarrow \infty$, each 
individual solution $\Psi_{q}(x_{j},t)$ behaves asymptotically as 
(\ref{e11}), implying that $\mathscr{C}(\Psi)\rightarrow C(\psi)$ in the asymptotic regime.

\textit{Confluence of Entanglement and Interference.}---For the entangled quantum system described by the wave function (\ref{filt_eq10}), the interference contribution of the probability density, $|\Psi^{\pm}\left(a,b, t\right)|^2$, can be expressed as
\begin{equation}
    I_{AB} \left(a,b, t\right) = 2|\xi\eta|\sqrt{\rho_{A}\rho_{B}}\cos(\Delta\varphi + \Delta\theta),
\label{eq_22}
\end{equation}
where we have written $\Phi_{A}=\sqrt{\rho_{A}}e^{i\varphi_{A}}$, $\Phi_{B}=\sqrt{\rho_{B}}e^{i\varphi_{B}}$, $\xi=|\xi|e^{i\theta_{\xi}}$, $\eta=|\eta|e^{i\theta_{\eta}}$,  with: $\Delta\varphi=\varphi_{A}-\varphi_{B}$ and $\Delta\theta=\theta_{\xi}-\theta_{\eta}$, $\rho_{A} = |\Phi_{A}|^2$ and $\rho_{B} = |\Phi_{B}|^2$.
Using equations (\ref{eq_2}) and (\ref{eq_17}) into equation (\ref{eq_22}), we can write a direct relation between the correlation interference $I_{AB}$ and our transient concurrence $\mathscr{C}(\Psi)$, namely,
\begin{equation}
I_{AB} \left(a,b, t\right) = \mathscr{C}(\Psi)\cos(\Delta\varphi + \Delta\theta).
\label{interference}
\end{equation}
This connection demonstrates that interference is not only a key component of the probability density in entangled quantum systems, but also involves information about the transient entanglement signatures captured by the transient concurrence

\begin{figure*}
\centering
\includegraphics[width=6.8in,height=3.3in]{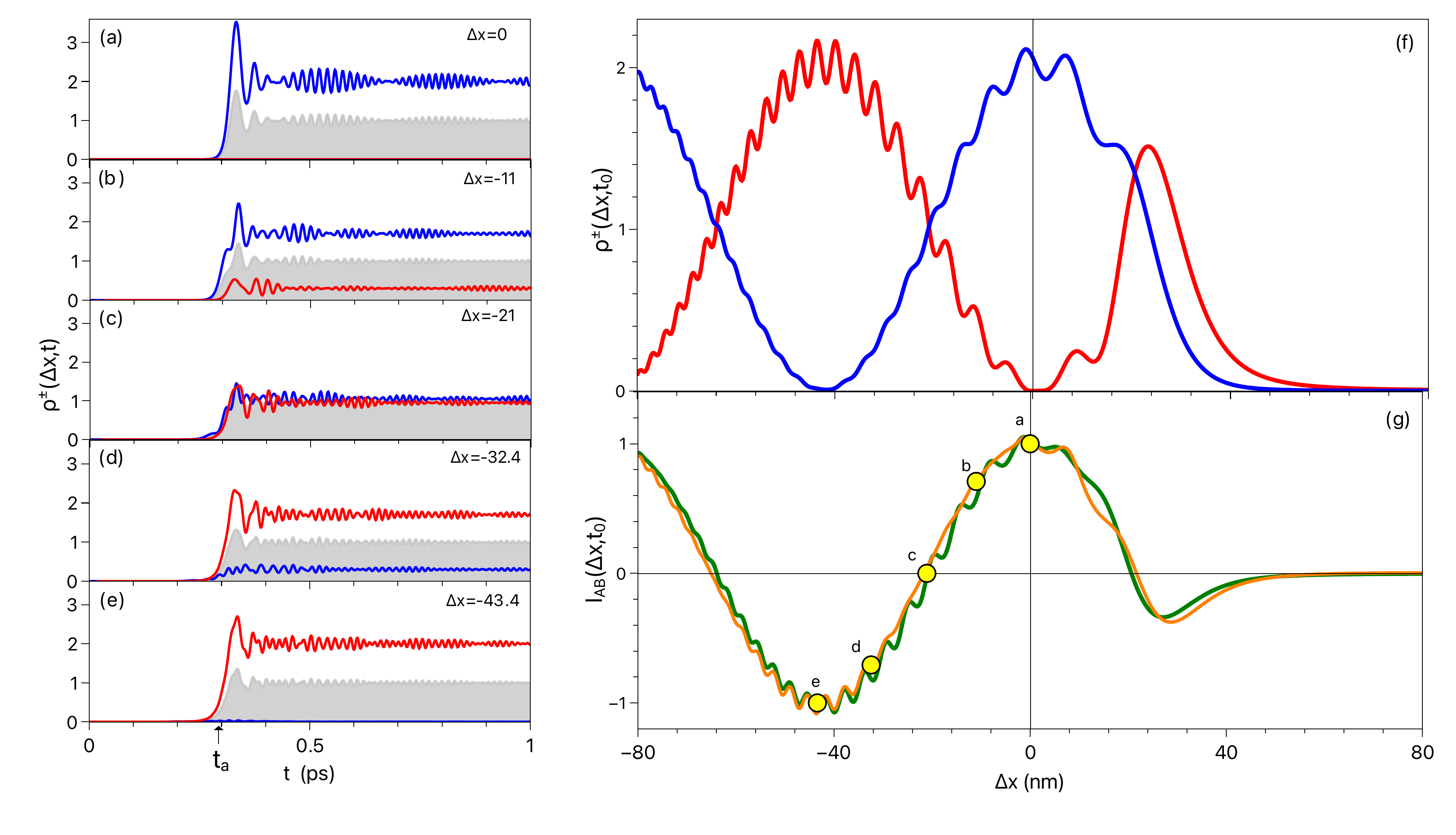}
\caption{\label{jointPD} On the left panel, we plot in (a) to (e) the symmetric (blue line) and antisymmetric (red line) joint probability densities as functions of time for the indicated fixed values of $\Delta x=b-a$ calculated from Eq. (\ref{Sol_Ms}). Parameters: $\xi_0=1 / \sqrt{2}$, $a = 100$ nm (varying $b$), $\alpha = 2.00 k_0$, $\beta = 2.05 k_0$ (the parameters $k_0$ and $m$ are the same as in Fig. \ref{Fig_1}). Also included here is the transient concurrence as a function of time (shaded gray curves). 
(f) Joint probability density as a function of $\Delta x$ at a fixed time $t_0=0.4$ ps, comparing the symmetric case (blue line) with the anti-symmetric case (red line).
(g) Correlation interference calculated with Eq. (\ref{interference_c}) (orange line) plotted as a function of $\Delta x$ compared to the exact calculation form Eq. (\ref{interference}) (green line).}
\end{figure*}

Figure \ref{jointPD} visualizes the relationship between the transient concurrence $\mathscr{C}(\Psi)$, the correlation interference $I_{AB}(\Delta x,t_0)$, and the time dependent joint probability density $\rho^{\pm}\left(\Delta x, t\right)=|\Psi^{ \pm}\left(a,b, t\right)|^2$, where $\Delta x=b-a$.
In the sequence of graphs on the left panel, we display the temporal evolution of $\rho^{\pm}\left(\Delta x, t\right)$ for fixed values of $\Delta x$, where $a=100$ nm (and choosing different values of $b$). Notice the strong fluctuations of $\rho^{\pm}$ as we follow from top to bottom the sequence of graphs for different values of $\Delta x$. We can clearly distinguish the bosonic from the fermionic cases since they fluctuate in opposite ways: when the plateau of $\rho^{+}$ reaches its maximum value, that of $\rho^{-}$ is minimum, and vice versa. 
As we can observe, for a given choice of detector positions $(a,b)$, the transient concurrence requires a certain activation time, $t_{\text{a}}$, before turning on. Prior to this time, we cannot distinguish between bosons and fermions through the values of $\rho^{\pm}\left(\Delta x, t\right)$. In our numerical example, we can see that this activation occurs at about $t_{\text{a}}\approx 0.3$ ps. The activation time $t_{\text{a}}$ can be defined quantitatively as the maximum propagation time required for all wavefront components to reach the detection points,
\begin{equation}
t_{\text{a}}=\max\left\{\frac{x_1}{v_\alpha},\frac{x_1}{v_\beta},
\frac{x_2}{v_\alpha},\frac{x_2}{v_\beta}\right\},
\label{activation}
\end{equation}
with $v_q=\hbar q/m$ ($q=\alpha,\beta$). 
For the parameters used in Fig.~2 this yields $t_a\simeq 0.297$ ps, 
in agreement with the onset observed in the numerical curves.

To provide the complementary perspective, on the right panel of Fig. \ref{jointPD}, we present graphs of the same quantities as functions of position difference $\Delta x$ at fixed time $t_0$. In the active region where $2|\xi \eta| \sqrt{\rho_A \rho_B} \ne 0$, $\rho^+$ and $\rho^-$ exhibit pronounced spatial oscillations corresponding to the fluctuations of the left panel graphs. Around the point $\Delta x=0$, we observe an accumulation (depletion) of the symmetric (anti-symmetric) joint probability density. Since $I_{AB}$ oscillates periodically, there are other values of $\Delta x$ at which the roles are reversed. For example, at the point marked with the letter $e$, $I_{AB}$ is negative, and consequently $\rho^{+}$ exhibits depletion while $\rho^{-}$ accumulation. This reversal is fully consistent with previous analyses showing that such inversions may arise from the spatial structure of the wavefunctions~\cite{MarchewkaGranot2014}. By continuously varying the detector positions, we move from $a$ to $e$ passing through the intermediate point $c$, where $I_{AB}=0$ and therefore $\rho^{+}=\rho^{-}$. 

The accumulation (or depletion) exhibited by the joint probability density is reminiscent of the bunching (or antibunching) behavior characteristic of 
HBT-type experiments. A continuous transition between antibunching and 
bunching has also been observed when detector positions are varied in recent 
HBT interferometry experiments~\cite{wolf2020}. Although the physical 
contexts and measured quantities are different, this parallel highlights the 
universality of the underlying quantum behavior. Bunching and antibunching have been successfully 
exploited as signatures of entanglement \cite{cano2011,baranger2013,Dumitrescu2017,thakkar2015}.

\textit{Tracing Quantum Connections: Signatures of bunching and anti-bunching.}---Let us explore these connections further. The Moshinsky function has the series expansion  \cite{ggcar97}
\begin{equation}
\begin{aligned}
M(y_{j,q}) = \frac{1}{2} e^{i m x_j^2/2 \hbar t}\left[2 e^{-i \gamma_{j,q}^{2}}+\frac{e^{i \pi / 4}}{\sqrt{\pi} \gamma_{j,q}}-\frac{e^{i^{3 \pi / 4}}}{\sqrt{\pi} \gamma_{j,q}^{3}}+\cdots\right],
\end{aligned}
\end{equation}
for $\pi/2 < arg(y_{j,q}) < 3 \pi/2$. If one adopts the exponential approximation 
$M(x_j, k_q, t) \approx \exp[i(m x_j^{2} / 2 \hbar t-\gamma_{j,q}^2)]$, the phase 
difference of $\Phi_A$ and $\Phi_B$ follows straightforwardly as
$\Delta\varphi=\gamma^2_{1,\beta} + \gamma^2_{2,\alpha}-\gamma^2_{1,\alpha} 
- \gamma^2_{2,\beta}$.
Since real coefficients $\xi$ and $\eta$ are often used, we can take $\Delta\theta=0$. Thus, after some simple algebra, we can derive an approximate expression for Eq. (\ref{interference}), namely
\begin{equation}
I_{AB} \approx \mathscr{C}(\Psi)\cos{(\Delta k \Delta x)},
\label{interference_c}
\end{equation}
where $\Delta k = \beta - \alpha$. The approximate formula shows excellent agreement with the exact calculation, as shown in Fig. \ref{jointPD} (g). Using this approximation, the time-dependent joint probability density takes the approximate form:

\begin{equation}
\begin{aligned}
\rho^{\pm}\left(\Delta x, t\right)\approx\xi^2 \rho_A+\eta^2 \rho_B 
\pm 2\xi\eta\sqrt{\rho_{A}\rho_{B}}\cos\left(\Delta k \Delta x\right).
\end{aligned}
\label{jpd_aprox}
\end{equation}
By progressing toward the stationary regime as $t \rightarrow \infty$, $\rho_A$ and $\rho_B$ approach unity according to Eqs. (\ref{e11}) and (\ref{eq_6}), and hence we have,
\begin{equation}
\rho^{ \pm} \left(\Delta x, t \rightarrow \infty\right) \approx 1 \pm C(\psi) \cos (\Delta k \Delta x),
\label{eq27}
\end{equation}
where $C(\psi)$ is the Wootters’s concurrence.

It is worth noting that, in the stationary regime, the extrema of 
$\rho^{\pm}$ are $\rho_{\max}=1+C(\psi)$ and 
$\rho_{\min}=1-C(\psi)$, so that the \emph{visibility}, a standard operational measure of interferometric contrast,
\begin{equation}
\mathcal V=\frac{\rho_{\max}-\rho_{\min}}{\rho_{\max}+\rho_{\min}}
\label{visibility}
\end{equation}
is equal to the Wootters concurrence $C(\psi)$. 
Thus, in this limit, the degree of entanglement directly determines the contrast
of the interference pattern.

In the case of maximal entanglement (i.e., $\xi = \eta = 1/\sqrt{2}$), we have $C(\psi)=1$ and hence we obtain the simplified and noteworthy result:  
\begin{equation}
\rho^{ \pm} \left(\Delta x, t \rightarrow \infty\right) \approx 1 \pm \cos (\Delta k \Delta x).
\label{asymtoticlimit}
\end{equation}

The latter expression has been a signature of the HBT effect as discussed and exemplified across varied physical systems, mainly in the context of quantum optics \cite{scully1997quantum2}. 
The comparison of (\ref{asymtoticlimit}) with the examples displayed in \cite{scully1997quantum2} and with recent analyses of stationary two-particle interference \cite{nazir2025} accentuates the ubiquity of the factor $1 \pm \cos(\Delta k \Delta x)$ as a universal indicator of quantum correlations across a diversity of quantum interference scenarios. 

Although our approach does not involve the statistical second-order correlation function, $g^{(2)}$, from Glauber’s quantum optics theory \cite{glauber1965, glauber2007quantum}, the oscillatory structure of our result mirrors the characteristic signatures of the Hanbury-Brown and Twiss effect, establishing a conceptual bridge between the dynamical manifestation of entanglement and interference phenomena reminiscent of HBT. The generalized stationary expression (\ref{eq27}) makes this relation explicit: concurrence modulates the characteristic oscillatory pattern.

The behavior of $\rho^{\pm}$, remarkably similar to that of the second-order
correlation function $g^{(2)}$ familiar from quantum optics, highlights a formal structural parallel between both descriptions. This motivates the adoption of
the terms \emph{probabilistic bunching} and \emph{probabilistic antibunching} to
describe, respectively, the accumulation and depletion of the joint probability
density, thereby promoting a more uniform terminology across the different physical domains where related quantum phenomena arise. It is worth recalling here that, as emphasized by
Lewenstein~\cite{lewenstein2007}, the quantum-optical understanding of photon
bunching originated from the interpretation of two-photon probability
amplitudes. This perspective played a central role in motivating Glauber’s
formulation of modern photon-counting theory~\cite{glauber1965,
glauber2007quantum}, where bunching and antibunching are characterized through
$g^{(2)}$. With this unified nomenclature in mind, the results displayed in
Fig.~\ref{jointPD} can be effectively summarized and compared with their
photonic counterparts~\cite{fox2006quantum} in
Table~\ref{table:quantum_phenomena}, where the classification based on $g^{(2)}$
is set side by side with the criteria defined through the interference term
$I_{AB}$ for bosonic and fermionic matter waves.

\begin{table}[h]
\caption{\label{table:quantum_phenomena}%
Classification of probabilistic bunching and probabilistic anti-bunching for (massive) bosonic and fermionic particles, compared to the well-known classification of bunching and anti-bunching in photonic systems \cite{fox2006quantum}.}

\begin{tabular}{lccc}
\toprule
\textrm{}&
\textrm{\textbf{Bunching}}&
\textrm{\textbf{Non-bunching}}&
\textrm{\textbf{Anti-bunching}}\\
\midrule
\addlinespace[1mm] 
\textbf{Photonic} & $g^2(0) > 1$ & $g^2(0) = 1$ & $g^2(0) < 1$ \\
\textbf{Bosonic} & $I_{AB} > 0$ & $I_{AB} = 0$ & $I_{AB} < 0$ \\
\textbf{Fermionic} & $I_{AB} < 0$ & $I_{AB} = 0$ & $I_{AB} > 0$ \\
\bottomrule
\end{tabular}
\end{table}

\textit{Conclusions and Outlook.}--- 
In this work, we have investigated the transient dynamics of copropagating entangled bosons and fermions by extending the quantum shutter model to analyze how entanglement manifests itself in the spatiotemporal structure of the joint probability density. Our approach, based on exact analytical time-dependent solutions, captures the phenomenon of diffraction in time and sets the foundation for exploring quantum transients in dynamical regimes.

We have introduced the notion of transient concurrence as a dynamical indicator of entanglement and demonstrated how it modulates the interference correlation in the joint probability density, revealing the active regions where probabilistic bunching and antibunching phenomena emerge. Furthermore, we have derived analytical expressions that establish a direct connection between entanglement and the oscillatory structure associated with the Hanbury–Brown and Twiss effect
thereby providing a theoretical framework to explore the manifestation of entanglement in transient regimes. Notably, in the case of maximum entanglement, we show that the asymptotic expression for the joint probability density reproduces the characteristic form $1 \pm \cos(\Delta k \Delta x)$, which is known to emerge in various quantum interference scenarios related to HBT-type interference. This connection between an abstract quantity (entanglement) and interference patterns opens new avenues for experimental investigations, potentially offering new methods to detect and characterize entanglement in dynamical quantum systems.

Our results establish structural connections between quantum transients, entanglement, and quantum optics. These findings not only enhance our understanding of the transient evolution of entangled systems but also motivate further exploration of copropagation scenarios in confined media, where interferometric correlations may play a key role in the study of quantum coherence and information dynamics.

\section*{Acknowledgments}
M. Á. Terán-Cruz. acknowledges financial support through BECA-CONAHCYT, México.
M. Á. Terán-Cruz and R.R. thank Jorge Villavicencio for useful discussions and suggestions.

\bibliographystyle{elsarticle-num}
\bibliography{references}

\end{document}